\documentstyle[preprint,tighten,aps,epsf]{revtex}

\def\Frac#1#2{\frac{\displaystyle{#1}}{\displaystyle{#2}}}
\def\lsim{\raise0.3ex\hbox{$\;<$\kern-0.75em\raise-1.1ex\hbox{$\sim\;$}}}
\def\gsim{\raise0.3ex\hbox{$\;>$\kern-0.75em\raise-1.1ex\hbox{$\sim\;$}}}
\begin{document}
\preprint{SISSA/69/2000/EP}

\title{Kaon vs. Bottom: Where to look for a general MSSM?}
\author{A. Masiero, O. Vives}
\address{SISSA -- ISAS, Via Beirut 4, I-34013, Trieste, Italy and \\ INFN,
Sezione di Trieste, Trieste, Italy}
\maketitle 
\begin{abstract}
We analyze CP violation and Flavor Changing effects in a general Minimal 
Supersymmetric extension of the Standard Model with arbitrary non--universal
soft--breaking terms. We show that, in this conditions, large FCNC effects are 
naturally expected in the Kaon system, even in the absence of quark--squark 
flavor misalignment. 
On the other hand, the B system is only sensitive to new supersymmetric 
contributions if the non--universality implies, not only different soft term
for the three generations but also the presence of a quark--squark 
misalignment much larger that the corresponding CKM mixing. 
The only exception to this rule are processes where the chirality changing 
contributions proportional to $\tan \beta$ are leading (for instance 
$b \rightarrow s \gamma$).
\end{abstract}

CP violation and Flavor Changing Neutral Current (FCNC) experiments are 
a main arena where searches of new physics beyond the Standard Model (SM)
of electroweak interactions will take place in the near future. In these 
processes, SM contributions are small and hence new contributions are allowed 
to compete on equal grounds. In fact, the currently available experimental 
information allows us to set stringent bounds on any extension of the SM in 
the neighborhood of the electroweak scale. The so--called minimal 
supersymmetric extension of the SM (MSSM) constitutes one of the most 
interesting examples of this. 
Indeed, it is well--known that a general MSSM with completely arbitrary 
soft--breaking terms at the electroweak scale suffers the so--called 
supersymmetric flavor and CP problems. This means that such a model tends 
to over--produce flavor changing and CP violation effects and this is used 
to set very stringent bounds on the structure of the sfermion mass matrices
\cite{gabbiani}.
However, the size of FCNC and CP violation effects are strongly dependent 
on the structure of the supersymmetry soft breaking terms which finally define
a particular MSSM. In a recent work \cite{blind,CPcons}, we showed that in the 
absence of a new flavor structures in the soft breaking terms no sizeable 
contributions to CP violation observables, as $\varepsilon_K$, 
$\varepsilon^\prime/\varepsilon$ or hadronic $B^0$ CP asymmetries, can be 
expected. In contrast, the presence of non--universality in the trilinear
terms, expected in string theories \cite{string,typeI}, is already enough to 
generate large supersymmetric contributions
to $\varepsilon^\prime/\varepsilon$ \cite{murayama,kkv,d12}. Unfortunately, 
the large theoretical uncertainties in the SM prediction prevent us from 
using this observable to identify these new supersymmetric contributions. 

In this letter, we will address the problem of indirect SUSY discovery in the 
framework of a completely general MSSM (see our definition below). This 
analysis does not intend to 
exhaust every single possibility but rather to provide a general view of the
``natural'' SUSY effects. To our knowledge, such a study is so far absent in 
the literature. The common attitude towards this problem consists in a pure
phenomenological analysis through the Mass Insertion (MI) approximation 
\cite{MI,gabbiani}. However, we will show that the inclusion of very general
and reasonable theoretical requirements has a deep impact on the expectations
of these SUSY effects when compared with MI bounds.  
In particular, we will see that, contrary to common wisdom, while large 
supersymmetric effects in hadronic B CP asymmetries are possible only under
some (interesting) special conditions \cite{brhlik2}, in the kaon system
these are naturally expected in any model with non--universal soft--breaking 
terms. More precisely: i){\sl if the new SUSY flavor structure is given by 
non--universal flavor diagonal soft terms, then only the kaon system is 
sizeably affected in FCNC and CP violation processes 
\footnote{However, as already known \cite{blind,bsgamma}, observables with a 
dominant 
chirality changing part whose SUSY contribution is proportional to
$\tan \beta$ (for instance $b\rightarrow s \gamma$ and
$b\rightarrow s l^+ l^-$) escape this rule, even in
the absence of new flavor structure, for large $\tan \beta$}, ii) if SUSY
signals are to be seen in CP violating hadronic B decays, then it implies
the presence, not only of flavor diagonal non--universality, but also a rather 
large quark--squark flavor misalignment (larger than the corresponding CKM 
mixings)}. 

In this context, it is interesting to comment the new results for the
$B\rightarrow J/\psi K_S$ CP asymmetry recently obtained by BaBar 
and BELLE \cite{BBB}. A possible discrepancy with the SM 
prediction could be due either to a dominant new physics contribution in the 
kaon system (i.e. modifying the determination of $\varepsilon_K$) or mainly to 
a new effect in the $B^0$ CP asymmetry itself \cite{BBtheory}. Taking into 
account the above statements, in a generic MSSM with non--universal 
soft--breaking 
terms it is very likely to have a sizeable effect in $\varepsilon_K$ while
a contribution to the $B^0$ CP asymmetry is possible only under special
conditions and again usually associated to a large contribution in the kaon
system. Hence, in an general MSSM as defined below, a low value in the 
$B\rightarrow J/\psi K_S$ CP asymmetry is more likely to be due to a 
new SUSY contribution to $\varepsilon_K$.  

In first place, we define our {\bf MSSM} through a set of four general
conditions:
\begin{itemize}
\item {\bf Minimal particle content}: we consider the MSSM, the minimal 
supersymmetrization of the Standard Model with no additional particles 
from $M_W$ to $M_{GUT}$.
\item {\bf Arbitrary Soft--Breaking terms ${\cal{O}}(m_{3/2})$}: The 
supersymmetry soft--breaking terms as given at the scale $M_{GUT}$ have a
completely general flavor structure. However, all of them are of the order of
a single scale, $m_{3/2}$. 
\item {\bf Trilinear couplings originate from Yukawa couplings}: Although
trilinear couplings are a completely new flavor structure they are related
to the Yukawas in the usual way: $Y^A_{i j} = A_{i j}\cdot Y_{i j}$, with
all $A_{i j}\simeq {\cal{O}}(m_{3/2})$.
\item {\bf Gauge coupling unification at $M_{GUT}$}.
\end{itemize}
In this framework, to define our MSSM, all we have to do is to write the 
full set of soft--breaking terms. This model includes, in the quark sector, 
7 different structures of flavor, $M_{\tilde{Q}}^2$, $M_{\tilde{U}}^2$, 
$M_{\tilde{D}}^2$, $Y_d$, $Y_u$, $Y^A_d$ and $Y^A_u$.
However, these 7 matrices are not completely observable, only relative 
misalignments among them are physical. Hence, we have the freedom to 
choose a convenient basis. At the supersymmetry breaking scale, $M_{GUT}$, 
the natural basis, specially from the point of view of the SUSY breaking
theory, is the basis where all the squark mass 
matrices, $M_{\tilde{Q}}^2,M_{\tilde{U}}^2, M_{\tilde{D}}^2$, are diagonal. 
In this basis, the Yukawa matrices are, $ v_1\, Y_d = {K^{D_L}}^\dagger\cdot 
M_d\cdot K^{D_R}$ and $ v_2\, Y_u = {K^{D_L}}^\dagger\cdot K^\dagger\cdot 
M_u\cdot K^{U_R}$, with $M_d$ and $M_u$ diagonal quark mass matrices, $K$ the 
Cabibbo--Kobayashi--Maskawa (CKM) mixing matrix and $K^{D_L}$, $K^{U_R}$, 
$K^{D_R}$ unknown, completely general, $3 \times 3$ unitary matrices. 

In a basis independent language
the matrices $K^{D_L}$, $K^{D_R}$, $K^{U_L}$ and $K^{U_R}$ measure the flavor
misalignment among, $d_L$--$\tilde{Q}_L$, $d_R$--$\tilde{d}_R$, 
$u_L$--$\tilde{Q}_L$ and $u_R$--$\tilde{u}_R$ respectively. 
Unfortunately, we have not much information on the structure of these matrices.
All we know is that there is, indeed, a minimum degree of misalignment among 
$K^{D_L}$ and $K^{U_L}$, which necessarily implies that these two matrices can 
not be simultaneously diagonal. This misalignment is described by the CKM 
mixing matrix $K^{U_L}\cdot {K^{D_L}}^\dagger = K$, and we use this relation 
to fix one of these matrices. There remain then, three unknown 
matrices, $K^{D_L}$, $K^{D_R}$ and $K^{U_R}$ . From the 
phenomenological point of view, we could have two complementary pictures. 
If the source of flavor structure in the SUSY breaking sector is completely 
independent from the source of flavor in the Yukawa sector, these matrices are
completely arbitrary and large mixings are possible. However, if the flavor 
structure in the Yukawa sector and the
soft--breaking terms have a common origin or are somehow related they
can be expected to be close to the identity, or more exactly, close to the CKM 
mixing matrix which is the minimum degree of misalignment experimentally 
required. As we will see below these two scenarios have very different 
phenomenological consequences. Finally, the trilinear matrices, $Y^A_d$ and 
$Y^A_u$, are specified in this basis by the SUSY breaking theory as 
$Y^A_{i j} = A_{i j}\cdot Y_{i j}$.

Once we specify these 7 matrices, our MSSM model is fully defined at the
$M_{GUT}$ scale. However, experimentally measurable quantities involve the 
sfermion mass matrices at the electroweak scale. So, the next step is to use 
the MSSM Renormalization Group Equations (RGE) \cite{RGE,bertolini} to evolve 
these matrices down to the electroweak scale. Below the electroweak scale, 
it is more convenient to work in the SCKM basis where the same unitary 
transformation is applied to both quarks and 
squarks so that quark mass matrices are diagonalized. The main RGE effects 
from $M_{GUT}$ to $M_W$ are those associated with the gluino mass and third 
generation Yukawa couplings. Regarding squark mass matrices, it is well--known
that diagonal elements receive important RGE contributions proportional to 
gluino mass that dilute the mass eigenstate non--degeneracy, 
$m^2_{\tilde{D}_{A_i}} (M_W) \simeq c_A^i \cdot m_{\tilde{g}}^2 + 
m_{\tilde{D}_{A_i}}^2$ \cite{RGE,bertolini,CPcons}, with 
$c_L^{1,2}\simeq (7, 7)$, $c_L^3 \simeq (5.8, 5.0)$, 
$c_R^{1,2}\simeq (6.6,6.6)$ and $c_R^3 \simeq (6.6, 4.5)$ for 
$(\tan \beta=2, \tan \beta=40)$. In the limit of complete degeneracy of the 
initial diagonal values at $M_{GUT}$ (equivalent to the Constrained MSSM) 
off--diagonal elements after RGE running are necessarily proportional to 
masses and CKM elements and hence 
small. This means that any larger new contribution to off--diagonal elements, 
possibly due to a large quark--squark misalignment in the different $K^A$ 
matrices, must be, at the same time, proportional to the difference of mass 
eigenstates at $M_{GUT}$ (for an example see Ref.\cite{hall}). 
Then, in the SCKM basis the off--diagonal 
elements in the sfermion mass matrices will be given by 
$(K^{A} \cdot M_{\tilde{A}}^2 \cdot {K^{A}}^\dagger)_{i\neq j}$ up to 
smaller RGE corrections.
On the other hand, gaugino effects in the trilinear 
RGE are always proportional to the Yukawa matrices, not to the trilinear 
matrices themselves and so they are always diagonal to extremely good 
approximation in the SCKM basis. Once more, the off--diagonal elements will 
be approximately given by 
$(K^{A} \cdot Y^{A}_{u,d} \cdot {K^{A}}^\dagger)_{i\neq j}$. 

After RGE running, flavor--changing effects in the SCKM basis can be estimated 
by the insertion of flavor--off--diagonal components of the mass--squared 
matrices normalized by an average squark mass, the so--called mass insertions
(MI)\cite{MI,gabbiani}.
In first place, we will analyze the $LR$ MI. Due to the trilinear terms 
structure, the $LR$ sfermion matrices are always suppressed by 
$m_{q}/m_{\tilde{q}}$, with $m_q$ a quark mass and $m_{\tilde{q}}$ the 
average squark mass. In any case, this 
suppression is compulsory to avoid charge and color breaking and directions 
unbounded from below \cite{casas}. In particular, it is required that 
$(Y^A_d)_{i j} \lsim \sqrt{3} m_{\tilde{q}}\ max\{m_i, m_j\}/v_1 $, which
in turn implies that $(\delta^{d}_{LR})_{i\neq j} \lsim \sqrt{3}\ 
max\{m_i, m_j\}/m_{\tilde{q}}$. So we must impose, as model independent 
upper bounds,
\begin{eqnarray}
\label{LRcolor}
(\delta^{d}_{LR})_{1 2}&\ \lsim& \sqrt{3}\ \Frac{m_s}{m_{\tilde{q}}}\ 
\simeq\ 3.2 \times 10^{-4} \cdot \left(\Frac{500\ GeV}{m_{\tilde{q}}} \right)
\nonumber \\
(\delta^{d}_{LR})_{1 3}&\ \lsim& \sqrt{3}\ \Frac{m_b}{m_{\tilde{q}}}\ 
\simeq\ 0.01 \cdot \left(\Frac{500\ GeV}{m_{\tilde{q}}} \right)
\end{eqnarray}
where we take all quark masses evaluated at $M_Z$ \cite{mb}.
The phenomenological MI bounds given in Table~\ref{tab:MI bounds} are 
stronger only for $Im (\delta^{d}_{LR})_{1 2}$. Hence, this  
means that there is still room to saturate $\varepsilon^\prime/\varepsilon$ 
with $LR$ mass insertions.
However, these bounds imply that, under general circumstances, it is not 
possible to saturate simultaneously $\varepsilon_K$ and 
$\varepsilon^\prime/\varepsilon$ with a 
$|(\delta^{d}_{LR})_{1 2}| = 3 \times 10^{-3}$ as suggested in 
Refs~\cite{brhlik2,murayama}\footnote{Unless the universe lives in a 
metastable vacuum for cosmologically long interval times}.  

In any case, we must remember that these are only upper bounds and a
definite model in the framework of our four general
conditions gives rise to somewhat smaller flavor changing effects.  
This can be seen explicitly in a type I string inspired example
\cite{typeI,kkv}. In these models, we can write the trilinear couplings 
in matrix notation as, 
\begin{eqnarray}
Y^A_{d (u)} (M_{GUT}) = \left(\begin{array}{ccc}
a^Q_1 & 0 & 0 \\ 0 & a^Q_2  & 0 \\ 0 & 0 & a^Q_3 \end{array}
\right) \cdot Y_{d (u)}\ +\ Y_{d (u)} \cdot \left(\begin{array}{ccc}
a^{D (U)}_1 & 0 & 0 \\ 0 & a^{D (U)}_2  & 0 \\ 0 & 0 & a^{D (U)}_3 \end{array}
\right) 
\end{eqnarray}
As discussed above, gluino RGE effects are again diagonal in the SCKM basis 
and off-diagonal elements are basically given by the initial conditions.
So, using unitarity of $K^{D_L}$ and $K^{D_R}$ it is straight--forward to get,
\begin{eqnarray}
\label{LRMI}
(\delta^{d}_{LR})_{i\neq j}&\ =\ \Frac{1}{m^{2}_{\tilde{q}}}\ \Big( m_j\ 
(a^Q_2 -  a^Q_1)\ K^{D_L}_{i 2} {K^{D_L}_{j 2}}^*\ +\  m_j\ (a^Q_3 -  
a^Q_1) K^{D_L}_{i 3} {K^{D_L}_{j 3}}^* \nonumber \\
&\ \ +\  m_i\  (a^D_2 -  a^D_1)\ K^{D_R}_{i 2} {K^{D_R}_{j 2}}^*\ +\  m_i\ 
(a^D_3 -  a^D_1)\ K^{D_R}_{i 3} {K^{D_R}_{j 3}}^* \Big) 
\end{eqnarray}
In the kaon system, we can make a simple estimate neglecting $m_d$ and 
assuming $K^{D_L}\approx K$, which is the expected size if quark and
squark masses have a common flavor origin (case i in the above introductory
discussion),
\begin{eqnarray}
\label{LRKaon}
(\delta^{d}_{LR})_{1 2}&\ \simeq\ \Frac{m_s}{m_{\tilde{q}}}\ 
\Frac{(a^Q_2 -  a^Q_1)}{m_{\tilde{q}}}\ K_{1 2} {K_{2 2}}^*\ \simeq\ a\
4 \times 10^{-5} \cdot \left(\Frac{500\ GeV}{m_{\tilde{q}}} \right)
\end{eqnarray}  
$a$ is a constant typically between $0.1$ and $1$.
Comparing with the bounds on the MI in Table~\ref{tab:MI bounds} we can see 
that indeed this value could give a very sizeable contribution to 
$\varepsilon^\prime/\varepsilon$\cite{murayama,kkv}. It is important to notice 
that the phase of $(a^Q_2 -  a^Q_1)$ is actually unconstrained by electric
dipole moment (EDM) experiments as emphasized in \cite{kkv}. This result is 
very important: it means that even if the relative quark--squark flavor 
misalignment is absent, the presence of non--universal
flavor--diagonal trilinear terms is enough to generate large FCNC effects
in the Kaon system.  

Similarly, in the neutral $B$ system, $(\delta^{d}_{LR})_{1 3}$ contributes 
to the $B_d-\bar{B}_d$ mixing parameter, $\Delta M_{B_d}$.
However, in the common flavor origin scenario (case i), i.e. 
$K^{D_L}\approx K$
(a direct extension of Ref.~\cite{murayama} mechanism), we obtain,
\begin{eqnarray}
\label{LRbCKM}
(\delta^{d}_{LR})_{1 3}&\ \simeq\ \Frac{m_b}{m_{\tilde{q}}}\ 
\Frac{(a^Q_3 -  a^Q_1)}{m_{\tilde{q}}}\ K_{1 3} {K_{3 3}}^*\ \simeq\ a\
 4.8 \times 10^{-5} \cdot \left(\Frac{500\ GeV}{m_{\tilde{q}}} \right),
\end{eqnarray}
clearly too small to generate sizeable $\tilde{b}$--$\tilde{d}$
transitions, as can be seen comparing to the MI bounds in 
Table~\ref{tab:MI bounds}.

Larger effects are still possible in our second scenario where flavor
origin in the Yukawa sector and the soft--breaking sectors are unrelated
(case ii).
In this framework, a large mixing could be possible, and with a maximal value,
$|K^{D_L}_{1 3} {K^{D_L}_{3 3}}^*| = 1/2$, we would approach  
\begin{eqnarray}
\label{LRbgen}
(\delta^{d}_{LR})_{1 3}&\ \simeq\ \Frac{m_b}{m_{\tilde{q}}}\
\Frac{(a^Q_3 -  a^Q_1)}{m_{\tilde{q}}}\ \Frac{1}{2}\ \simeq\ a\
3 \times 10^{-3} \cdot \left(\Frac{500\ GeV}{m_{\tilde{q}}} \right).
\end{eqnarray}
Even in this limiting situation, this result is roughly one order of
magnitude too small to saturate $\Delta M_{B_d}$, though it could still be 
observed
through the CP asymmetries. Hence in the $B$ system we arrive to a very 
different result: it is not at all enough to have non--universal trilinear 
terms, but, it is also required to have large flavor
misalignment among quarks and squarks.   

The situation for the $LL$ and $RR$ mass insertions is less defined 
due to the absence of any theoretical prejudice on these mass matrices at 
$M_{GUT}$. In any case we can write these MI as,
\begin{eqnarray}
\label{AMI}
(\delta^{d}_{A})_{i\neq j}&\ =\ \Frac{1}{m^{2}_{\tilde{q}}}\ \Big(
(m_{\tilde{D}_{A_2}}^2 - m_{\tilde{D}_{A_1}}^2 )\ K^{D_A}_{i 2} {K^{D_A}_{j
2}}^*\  +\ (m_{\tilde{D}_{A_3}}^2 - m_{\tilde{D}_{A_1}}^2 )\ K^{D_A}_{i 3} 
{K^{D_A}_{j 3}}^* \Big) 
\end{eqnarray}
However, due to the gluino dominance in the squark eigenstates at $M_W$ we 
can say that $ m^{2}_{\tilde{q}}(M_W)\approx 7\ m_{\tilde{g}}^2(M_{GUT})$. 
Hence, $(m_{\tilde{D}_{A_i}}^2 - m_{\tilde{D}_{A_1}}^2)/m^{2}_{\tilde{q}} 
\approx 0.1\ b \ m^{2}_{\tilde{q}}$ with $b$ a number ${\cal{O}}(1)$.
Replacing these values in Eq~(\ref{AMI}), this means, for the kaon system and
assuming CKM--like mixing (case i),
\begin{eqnarray}
\label{dL12}
(\delta^{d}_{A})_{1 2}&\ \simeq\ 0.1\ b\ K_{1 2}\simeq 0.02\ b. 
\end{eqnarray}
Hence, in the presence of non--universality, sizeable SUSY
contributions to the kaon mass difference are expected. Regarding CP
violation observables, with large phases in the $K^{D_A}$ matrices, 
big contributions to $\varepsilon_K$ arise (notice $m_{\tilde{D}_{A_i}}^2$
are always real numbers). This is again equivalent to our result for the 
$LR$ transitions: even if the sfermion mass matrices at $M_{GUT}$ are 
flavor diagonal, large FCNC effects are produced in the $K$ system with
family non--universal masses.

In the B system, we must again distinguish our two possible
scenarios. In case i), $K^{D_L}\approx K^{D_R}
\approx K$,
\begin{eqnarray} 
\label{dL13}
(\delta^{d}_{A})_{1 3}&\ \simeq\ 0.1\ b\ K_{1 3}\simeq 8 \times 10^{-4}\ b, 
\end{eqnarray}
clearly too small to be observed at the B factories. Nevertheless, in
the presence of large mixing, still possible when flavor in the soft
breaking terms has an independent source, we can get $K^{D_A}_{1 3}
{K^{D_A}_{3 3}}^* \simeq 1/2$. In this limiting situation 
$(\delta^{d}_{A})_{1 3} \lsim 0.05$, still reachable through CP
asymmetries at the B factories in the presence of a sizeable phase in 
$K^{D_A}$. So, we find again that to have observable effects in the $B$ system
it is required to have not only the presence of non--universality but also 
large quark--squark flavor misalignment.  

In summary, in gluino mediated transitions large effects
are expected in the Kaon system in the presence of non--universal squark masses
even with a ``natural'' CKM--like mixing. 
However in the $B$ system, due to the much lower sensitivity to supersymmetric 
contributions, observable effects are expected only with approximately maximal 
$\tilde{b}$--$\tilde{d}$ mixings.

Finally we add a short comment on chargino contributions. The chargino sector 
is more involved due to the presence of the CKM mixing
matrix and the additional mixing wino--higgsino. However, for large and 
similar squark mixings in the up and down sector, chargino and gluino 
contributions are expected to be of the same order, barring special situations 
as light stop ($M_{\tilde{t}} \lsim 300\ GeV$) and chargino. 
Indeed, in this case, the effects of the CKM matrix in chargino couplings are 
sufficiently small and can be neglected. Then, the flavor changing chargino 
and gluino vertices have similar mixing matrices and the difference in the 
couplings is compensated through the RGE relation $\alpha_3/m^2_{\tilde{g}} = 
\alpha_2/m^2_{\tilde{w}}$.
The only additional ingredient in chargino vertices is the presence of 
higgsino. Higgsino couplings are suppressed by small Yukawa couplings
with the exception of the top quark with $Y_t \approx 1$.
Hence, the main difference between gluino and chargino contributions is the 
possible presence of a light stop due to the the large top quark Yukawa 
coupling.   
The analysis of these new SUSY flavor contributions to $B_d$--$\bar{B}_d$ 
mixing in the presence of light stop and chargino was considered in 
Ref~\cite{brhlik2}\footnote{For a discussion in the flavor universal case see
\cite{Buniversal}}. 
It was shown there that for a light stop of $140\ GeV$ and a chargino of 
$100\ GeV$ and with a large $\tilde{u}_L$--$\tilde{t}_L$ mixing, 
$K^{U_L}_{1 3} \simeq \lambda_c$ it is still possible to produce contributions 
to $\sin 2 \beta$ as large as $0.78$ with $\delta_{CKM}=0$. However, this is 
a limiting case, light stop and chargino and, more important, a sizeable 
mixing in the ${\tilde{u}_L}$--${\tilde{t}_L}$ are required.
Hence, we conclude that also chargino contributions fit on our general 
scheme and only in an scenario with different flavor origin in the Yukawa
and soft--breaking sectors sizeable effects appear.

In conclusion we have shown that in the kaon system large SUSY effects are 
naturally expected in any model with non--universal soft--breaking terms.
In this direction, several works can be found in the literature 
\cite{buras-colang,buras}
in the mass insertion context, and a complete analysis will be presented in 
short time \cite{WIP}. 
On the other hand, a discovery of a supersymmetric contribution to hadronic 
$B^0$ CP asymmetries would signal the presence of a different origin of flavor 
in the breaking of SUSY.
  
We thank T. Kobayashi, E. Lunghi, M. Misiak and J.C. Pati for enlightening 
discussions.
The work of A.M. was partially supported by the European TMR Project
``Beyond the Standard Model'' contract N. ERBFMRX CT96 0090; O.V. 
acknowledges financial support from a Marie Curie EC grant 
(TMR-ERBFMBI CT98 3087) and partial support from spanish CICYT 
AEN-99/0692. 

%\begin{references}

%\end{references}
\begin{table}
  \caption{Mass insertion bounds with $x\simeq1$ and 
    $m_{\tilde{q}} = 500\ GeV$. Coming from gluino contributions, they are 
   completely simmetric under the exchange $L L\leftrightarrow R R$, or
    $L R\leftrightarrow R L$}
  \label{tab:MI bounds}
  \begin{tabular}{cccc}
    %\hline 
    \multicolumn{2}{c}{$\Delta m_{K}$}&\multicolumn{2}{c}{$\varepsilon_K$} \\
    \noalign{\smallskip} \hline \noalign{\smallskip}
    $|{\rm Re}(\delta^{d}_{LL})_{12}^{2}|^{\frac{1}{2}}$
    & $|{\rm Re}(\delta^{d}_{LR})_{12}^{2}|^{\frac{1}{2}}$
    & $|{\rm Im}(\delta^{d}_{LL})_{12}^{2}|^{\frac{1}{2}}$
    & $|{\rm Im}(\delta^{d}_{LR})_{12}^{2}|^{\frac{1}{2}}$\\
    \noalign{\smallskip} \hline \noalign{\smallskip}
    0.040 & 0.0044 &  0.0032 & 0.00035\\
    \noalign{\smallskip}\hline \hline \noalign{\smallskip}
    \multicolumn{2}{c}{$\varepsilon^\prime/\varepsilon$}&
    \multicolumn{2}{c}{$\Delta m_{B_d}$} \\ 
    \noalign{\smallskip} \hline \noalign{\smallskip}
    ${\rm Im}(\delta^{d}_{LL})_{12}$
    & ${\rm Im}(\delta^{d}_{LR})_{12}$
    & $|{\rm Re}(\delta^{d}_{LL})_{13}^{2}|^{\frac{1}{2}}$
    & $|{\rm Re}(\delta^{d}_{LR})_{13}^{2}|^{\frac{1}{2}}$\\
    \noalign{\smallskip} \hline \noalign{\smallskip}
    0.50 & 0.000021 & 0.098 & 0.033 \\
    \noalign{\smallskip}%\hline
  \end{tabular}
\end{table}

\end{document}